# Superconductivity in the intermetallic compound YRhAl

Darshan C. Kundaliya[†] and S.K. Malik*
*Tata Institute of Fundamental Research, Colaba, Mumbai, 400 005, India*

**Abstract**

The intermetallic compound, YRhAl, has been prepared and is found to be isomorphic with RRhAl (R=Pr, Nd, Gd, Ho and Tm) compounds crystallizing in the orthorhombic TiNiSi-type structure (space group *Pnma*). Heat capacity and electrical resistivity measurements in the He-3 temperature range reveal that this compound is superconducting with a transition temperature, $T_c$, of 0.9K. The electronic specific heat coefficient, $\gamma$, and the Debye temperature are found to be 6.1mJ/mole K and 197 K, respectively. The specific heat jump at the superconducting transition is found to be consistent with the BCS weak-coupling limit. This combined with the earlier observation of superconductivity in LaRhAl ($T_c$=2.4K) having a different structure then that of YRhAl, suggests that the underlying structure is not very crucial for the occurrence of superconductivity in RRhAl series of compounds.

*PACS*: 71.20.Lp; 74.25 Ha; 74.25.Bt; 74.25.-q

*Keywords:* A. Intermetallic compound; D. Heat capacity; A. Superconductivity; D. Resistivity

*Corresponding author: Email: skm@tifr.res.in
Fax: +91-22-2280 4610
Phone: +91-22-2280 4545

# 1. Introduction

In the recent past, there has been a great deal of interest in the study of electronic magnetic and transport properties of equiatomic ternary rare earth intermetallic compounds of the type RTX, where R is a rare earth, T is a transition metal element and X is an *s-p* element. In particular, the properties of the Ce- and the La-based compounds have been widely studied. The Ce compounds often show a variety of ground states depending upon the hybridization of the Ce-4f electrons with the conduction electrons. The corresponding La compounds are of interest for comparison since La lacks any 4f electrons. In our laboratory, we observed a Kondo insulating ground state for CeRhSb [1] and a superconducting ground state in its La analog [2]. To our knowledge, the RRhSb series is the only series in which the Ce-based compound is an insulator while the La-based compound is superconducting. Mixed valent behaviour was observed in CeIrGa [3], CeIrAl [4], and CeRhAl [5] compounds while superconductivity was observed in LaRhAl [5]. Kondo lattice behaviour was observed in CeNiSb [6] and CeAuAl [7].

The RRhAl (R=rare earth) compounds form in two different structure types [8]. Those with R=La and Ce form in the orthorhombic $Pd_2(MnPd)Ge_2$ type structure while those with R=Pr, Nd, Gd, Ho and Tm form in the orthorhombic TiNiSi-type structure. Superconductivity, with a transition temperature of 2.4K, was observed by our group in LaRhAl forming in the $Pd_2(MnPd)Ge_2$ structure. Therefore, it was of interest to see whether other nonmagnetic compounds of the RRhAl series, forming in the TiNiSi-type structure, will also exhibit superconductivity. With this in mind, we have now prepared and examined another member of this series, namely, YRhAl. We find that this compound also adopts the TiNiSi-type structure - same as that of the RRhAl compounds with R=Pr, Nd, Gd, Ho and Tm. Further, this compound is found to be superconducting with a transition temperature of ~0.9 K. In this communication we present the results of low temperature heat capacity and electrical resistivity measurements on this compound.

## 2. Experimental Details

The compound YRhAl was prepared by melting together stoichiometric amounts of the constituent elements in an arc furnace on a water-cooled copper hearth under a continuous flow of argon gas. The starting materials were at least 99.9% pure. The alloy button was turned over and melted several times to ensure homogenous mixing. The phase purity of the sample was checked by powder x-ray diffraction using CuK$_\alpha$ radiation on a Siemens X-ray diffractometer. Heat capacity measurements, in the temperature range of 0.37 K to 10 K and in various applied fields, were carried out using the relaxation method (PPMS, Quantum Design). Four-probe dc electrical resistivity measurements were also carried out using the same system in the temperature range of 0.37-300 K and in various applied fields. Magnetization measurements were made in the temperature range of 1.8-300 K using a Squid Magnetometer (MPMS, Quantum Design).

## 3. Results and Discussions

### 3.1. Structure

Powder x-ray diffraction studies revealed that YRhAl is a single-phase compound; isomorphous with other RRhAl (R=Pr, Nd, Gd, Ho and Tm) compounds crystallizing in the orthorhombic TiNiSi structure (space group *Pnma*, *No. 62*). The lattice parameters, obtained from a Rietveld fit of the x-ray data based on the above structure, are:

a=6.8535(7) Å, b=4.1786(4) Å and c=7.9286(8) Å.

These lattice parameters are consistent with the lattice parameters observed in other RRhAl compounds with R=Gd and Ho (Ref. 8). In contrast to the crystal structure of YRhAl, that of LaRhAl is a kind of mixture of TiNiSi type and ZrNiAl type and has almost double the c lattice parameter of YRhAl [8]. The coordination of Y and La as well as that of Al is quite similar in the two structures. However, the coordination of Rh in YRhAl is an average of the two Rh sites in LaRhAl.

### 3.2. Magnetic Susceptibility

The magnetic susceptibility of YRhAl, in the temperature range of 1.8-300 K, is shown in Fig. 1. The compound shows a weakly temperature dependent susceptibility in the temperature range of 300-20 K. A rise in susceptibility is seen at temperatures below ~20 K which is possibly due to traces of magnetic impurities in the starting materials. In fact, a modified Curie-Weiss fit to the susceptibility with the equation $\chi=\chi_0+C/(T-\theta_p)$, yields an effective magnetic moment of $0.1\mu_B$. The temperature independent susceptibility, $\chi_0$, is $8.0\times10^{-5}$ emu/mole which is close to the value of the total susceptibility of $9.2\times10^{-5}$ emu/mole near room temperature. Superconductivity in this compound is not observed through magnetization measurements down to 1.8 K - the lowest attainable in the Squid magnetometer.

### 3.3. Electrical Resistivity

Figure 2 shows a plot of electrical resistivity versus temperature for YRhAl in the temperature range of 0.37-300 K. The resistivity of this compound shows metallic behaviour in the above temperature range. A transition to the superconducting state is observed starting at about 0.9 K with zero resistance reaching at about 0.7 K. The effect of magnetic field on the resistivity near the superconducting transition is shown as an inset in Fig. 2. The superconducting transition temperature is rapidly suppressed by the application of even a small magnetic field to the extent that the transition is pushed below 0.37 K (the lowest temperature in our He-3 set up) in a field of ~500 Oe.

### *3.4. Heat Capacity*

Figure 3 shows the plot of heat capacity C versus temperature T for YRhAl in zero applied field as well as in an applied magnetic field of 1 Tesla. A jump in the heat capacity is observed starting at ~0.9 K indicative of the bulk superconducting transition. This jump is suppressed by the application of a magnetic field of 1 T. The low temperature specific heat data are fitted to the equation $C=\gamma T+\beta T^3$, where $\gamma$ is the electronic specific heat coefficient and $\beta T^3$ represents the lattice contribution. Figure 4 shows a plot of C/T vs. $T^2$ for YRhAl. Fit to the heat capacity data by the above equation is also shown in this figure and yields $\gamma=6.1\pm0.1$ mJ/mole K and $\beta=0.2536\pm0.0013$. The $\gamma$ value of YRhAl is typical of metallic systems. The Debye temperature, $\Theta_D$, is estimated from the classical expression $\Theta_D=(12\pi^4 nR/5\beta)^{1/3}$, where n is the number of atoms per formula unit and R is the gas constant. From the value of $\beta$ mentioned above, $\Theta_D$ is found to be 197±1 K Both, the values of $\gamma$ and $\Theta_D$ are comparable to corresponding values of 7.7±0.1 mJ/mole K and 252±3K found for the ternary compound LaRhSb [2] isostructural with YRhAl. The smaller value of $\Theta_D$ for YRhAl points towards a relatively softer lattice and hence a lower superconducting transition temperature.

The superconducting transition is somewhat rounded and, therefore, the value of the jump, $\Delta C$, in the specific heat at the superconducting transition cannot be exactly delineated. If the C vs T curve in the superconducting state is extrapolated to 0.9 K, where the onset of superconductivity occurs, then the jump ΔC is estimated to be 5.93 mJ/mole K. The ratio $\Delta C/\gamma T_c$ is found to have a value of 1.08 which may be compared with the BCS weak coupling limit of 1.43 [9]. This value suggests that YRhAl is a typical BCS weak coupling superconductor. Observation of superconductivity in YRhAl along with that in LaRhAl with two different crystal structures suggests that superconductivity in this series of compounds is not sensitive to the underlying structure type. The transition temperature, however, depends on the unit cell volume mimicking the effect of pressure.

## 4. Conclusions

In conclusion, a new compound, YRhAl, has been prepared and is found to be isomorphic with other RRhAl (R=Pr, Nd, Gd, Ho and Tm) compounds. Low temperature heat capacity and electrical resistivity measurements show that this compound is a superconductor with transition temperature of 0.9 K. The electronic specific heat coefficient is found to be 6.1 mJ/mole K. The specific heat jump at superconducting transition temperature is consistent with BCS weak coupling limit. Observation of superconductivity in LaRhAl and YRhAl – two members of the same series but crystallizing in two different structures - suggests that the underlying structure in this series of compounds is not significant for the occurrence of superconductivity.

**Figure Captions**

Figure 1. Magnetic susceptibility ($\chi$) versus temperature (T) and $\chi^{-1}$ versus T for YRhAl.

Figure 2. Electrical resistivity versus temperature for YRhAl. The inset shows the temperature variation of resistivity near the superconducting transition temperature in the presence of various applied fields.

Figure 3. Heat capacity (C) versus temperature (T) in zero and 1 Tesla applied magnetic field for YRhAl.

Figure 4. Heat capacity (C) as a function of temperature (T) plotted as C/T vs $T^2$ in zero and 1 Tesla applied magnetic field for YRhAl.

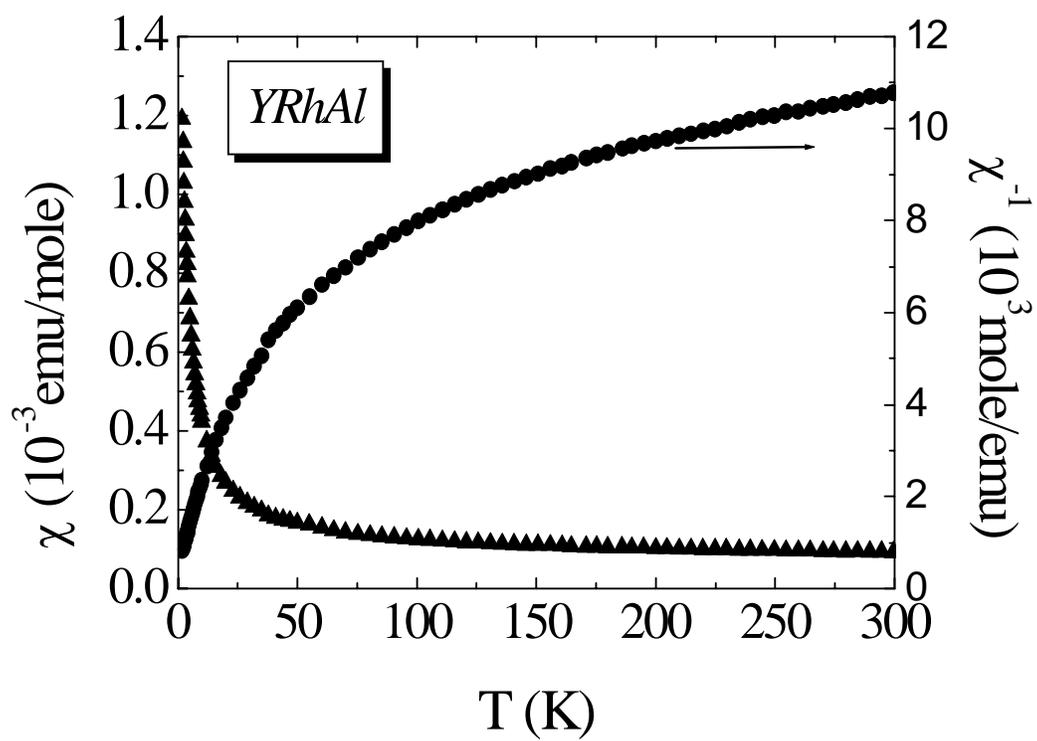

Figure 1 *Darshan et al.*

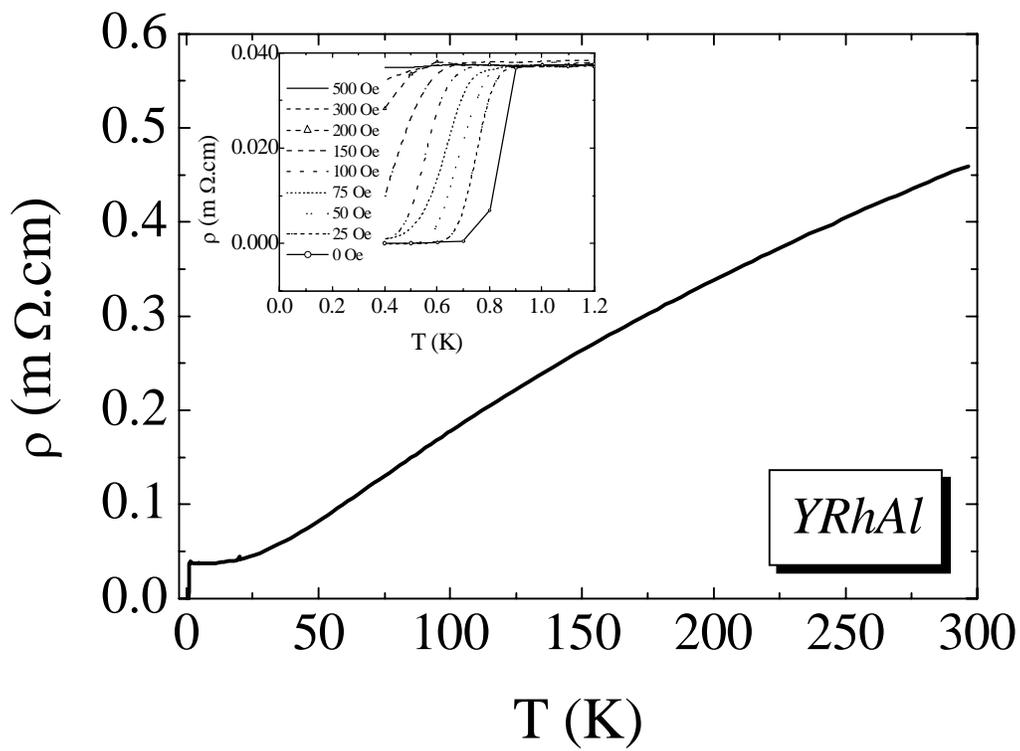

Figure 2 *Darshan et al.*

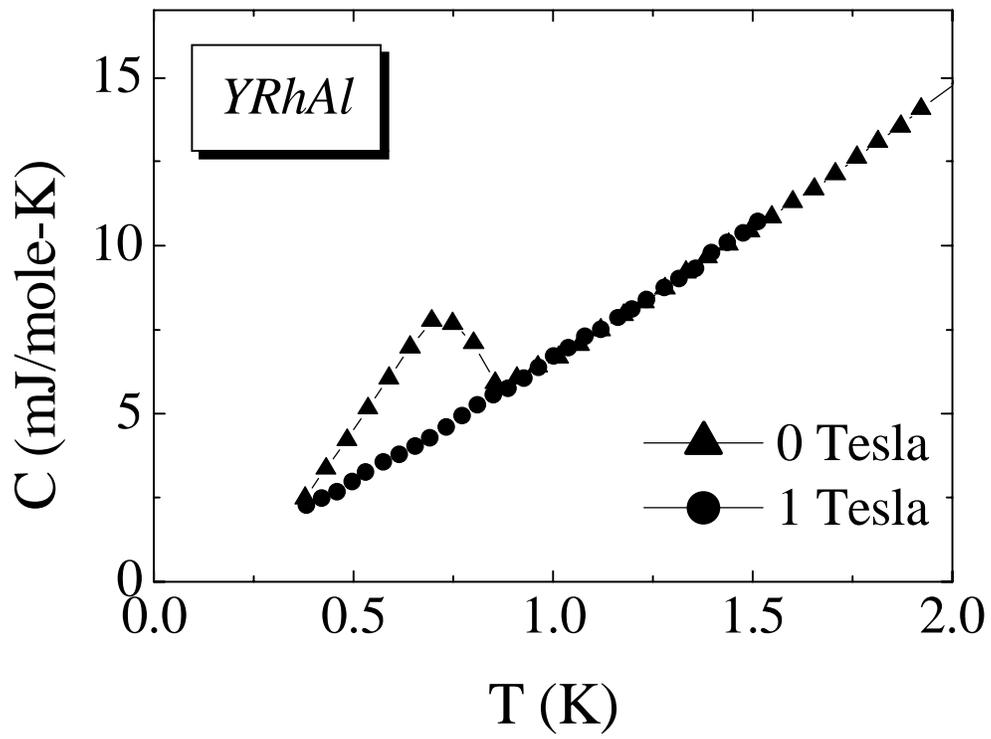

Figure 3 *Darshan et al.*

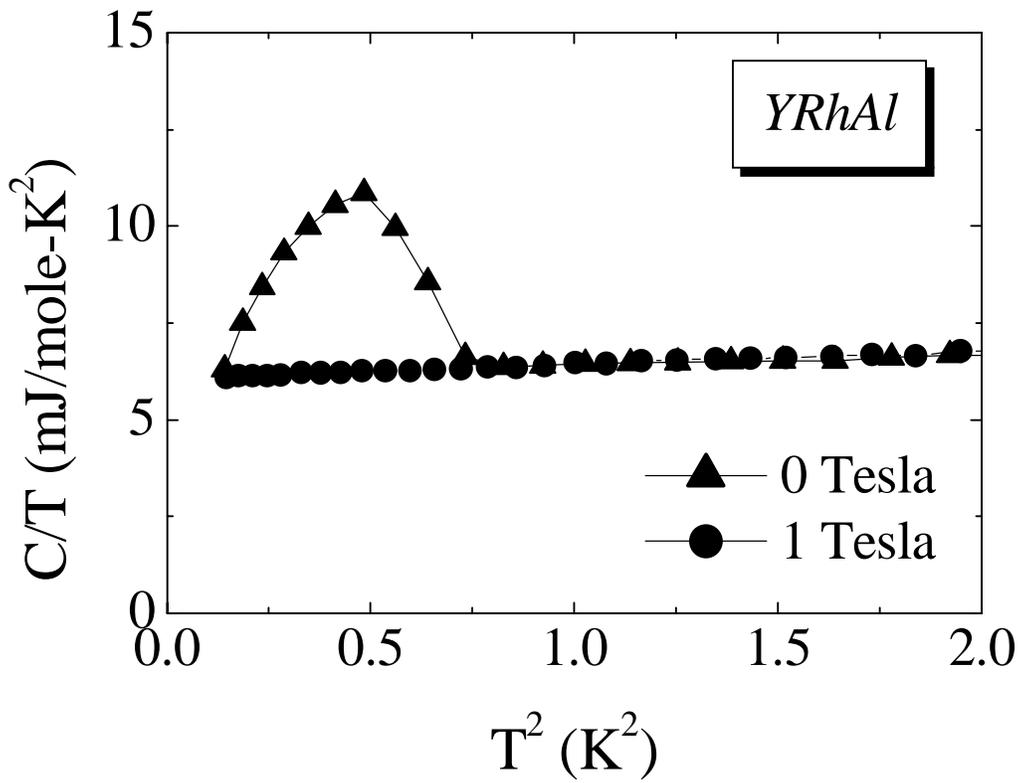

Figure 4 *Darshan et al.*